\documentclass[prd]{revtex4}

\usepackage{graphicx}
\usepackage{dcolumn}
\usepackage{bm}

\begin{document}

\title{Embedding spherical spacelike slices in a Schwarzschild solution}
\author{Niall \'O Murchadha}
\email{niall@ucc.ie}
\affiliation{Physics Department, University College, Cork, Ireland}
 \affiliation{ESI, A-1090 Wien, Boltzmanngasse 9, Austria}
\author{Krzysztof Roszkowski}%
 \email{roszkows@if.uj.edu.pl}
\affiliation{Institute of Physics, Jagellonian University, Krak\'ow,\ Poland}%
\affiliation{Physics Department, University College, Cork, Ireland}
 \affiliation{ESI, A-1090 Wien, Boltzmanngasse 9, Austria}

\date{\today}

\begin{abstract}
Given a spherical spacelike three-geometry, there exists a very simple algebraic condition which tells us whether, and in which, Schwarzschild solution this geometry can be smoothly embedded. One can use this result to show that any given Schwarzschild solution covers a significant subset of spherical superspace and these subsets form a sequence of nested domains as the Schwarzschild mass increases. This also demonstrates that spherical data offer an immediate counter example to the thick sandwich `theorem'.
\end{abstract}

\maketitle
\section{Spherical spacelike slices in a Schwarzschild spacetime}
It is clear that many spherical spacelike slices can be embedded in a given extended Schwarzschild spacetime. In the negative mass Schwarzschild (or in flat space-time) all the slices will have topology $R^3$. In the positive mass Schwarzschild one has a range of topologies. There are slices which start at one spacelike infinity, run through the middle and out the other end. These have topology $S^2 \times R^1$. There are slices which start at one or other of the $R = 0$ singularities and go out to one or other of the infinities. These have topology $R^3$. Finally, there are slices which start and end on one or other of the singularities. These have topology $S^3$. There are variants of the infinite ones, where instead of going to spacelike infinity, the slice remains within the horizon and runs more or less along one of the $R$ = constant lines, with $R < 2m$. These slices are asymptotically cylindrical. Inside the horizons, the lines of constant $R$ are spacelike, while outside they are timelike. Therefore inside the horizons a spacelike spherical slice can have an oscillatory areal radius with as many maxima and minima as one wishes, while outside the horizon the area must monotonically increase. These features are, of course, reflected in the nature of the spacetime in which any given slice can be embedded. 

Any spacelike slice embedded in a solution of the Einstein equations will have an intrinsic geometry, given by a three-metric $g_{ab}$, and an extrinsic curvature, given by a symmetric tensor $K^{ab}$. These are not independent, they must satisfy the constraints 
\begin{equation} {\cal R}^{(3)} - K_{ab}K^{ab} + \left(g_{ab}K^{ab}\right)^2   = 0, \label{ham} \end{equation} \begin{equation} \nabla_a\left(K^{ab} - g^{ab}g_{cd}K^{cd}\right) = 0, \label{mom} \end{equation} 
where ${\cal R}^{(3)}$ is the scalar curvature of $g_{ab}$. Eq.(\ref{ham}) is called the Hamiltonian constraint, Eq.(\ref{mom}) is called the momentum constraint.

 Given a spherical slice in a spherical space-time then both the three-geometry and the extrinsic curvature are spherically symmetric. There are several `natural' choices of coordinates that are used to write down the spherical three-metric. The one we favor is the `proper distance gauge' where one writes the three-metric as \begin{equation} ds^2 = dl^2 + R^2(l) d\Omega^2, \label{g} \end{equation} where $d\Omega^2 = d\theta^2 + \sin ^2\theta d \phi^2$ is the round two-metric on a sphere, $R(l)$ is the areal (Schwarzschild) radius of the isometry two-spheres (expressed as a function of $l$) and $l$ is the proper distance between surfaces of constant $R$. Any spherical three-geometry can be expressed in this form. If the topology of the slice is $S^2 \times R$, $l \in (-\infty, \infty)$, if the topology is $R^3$, $l\in [0, \infty)$, and if the topology is $S^3$, $l\in [-a, +b]$ where $a + b$ is the proper distance between the points where $R \rightarrow 0$. 

A spherically-symmetric symmetric two-tensor has only two independent components. We can write the extrinsic curvature in a form consistent with spherical symmetry as \cite{gom} 
\begin{equation} \label{k} K^{ab} = n^an^bK_l + (g^{ab} - n^an^b)K_R, \end{equation} 
where $K_l$ and $K_R$ are two scalars and $n^a$ is the outward-pointing unit normal to the two-surfaces of constant $R$, $n^a = (1, 0, 0)$ in our preferred gauge. 

The Hamiltonian and momentum constraints can now be written in the special case of spherical symmetry as \cite{gom} 
\begin{equation} K_R\left[K_R + 2K_l\right] - {1 \over R^2}\left[ 2RR'' + R'^2 - 1\right] = 0, \label{hs} \end{equation}
 and 
\begin{equation} K'_R +{R' \over R}\left[K_R - K_l\right] = 0, \label{ms} \end{equation} 
where $'$ represents the derivative with respect to $l$. There is a first integral of the constraints \cite{gom} 
\begin{equation} m = {R^3K^2_R \over 2} + {R \over 2}\left[1 - R'^2\right], \label{m} \end{equation} 
where $m$ is the Schwarzschild mass. This is the Misner-Sharp, Hawking {\it et al} mass formula. An immediate consequence of Eq.(\ref{m}) is that if we define \begin{equation} \label{M} M = \max{R \over 2}\left[1 - R'^2\right], \end{equation} 
where the maximum is taken over the whole three-geometry then 
\begin{equation} \label{m>M} m \ge M. \end{equation} 
This is our key equation, it is an algebraic relation between a global spacetime quantity, $m$, and a quantity, $M$, which depends only on the three-geometry. 

In this article we will show that essentially the converse of condition (\ref{m>M}) holds. More precisely, given a three-geometry in the proper distance gauge, one can always compute the quantity $M$ as defined by Eq.(\ref{M}). Let us assume that it is finite. This only requires that $R \sim l$ as $l \rightarrow \pm \infty$. We will show that this spacelike spherical geometry can be embedded in any Schwarzschild solution whose Schwarzschild mass satisfies \begin{equation} m > M. \label{M<m} \end{equation} The quantity $M$ is really a coordinate independent object. $R$ is the area of the two-dimensional isometry spheres and $2R'/R$ is the mean extrinsic curvature of the isometry two-spheres as embedded surfaces in the three-geometry.

\section{Embedding slices in a Schwarzschild solution} 
Let us assume that we are given a spherical three metric, say in the form of Eq.(\ref{g}). It could be a complete manifold or just a patch. We wish to consider this as the metric of a hypersurface in a spherical vacuum solution of the Einstein equations. Therefore we need to find a spherical extrinsic curvature $(K^{ab})$ expressed, say, in the form of Eq.(\ref{k}), such that the combination $(g_{ab}, K^{ab})$ satisfy both the Hamiltonian and momentum constraints, Eqs. (\ref{ham}) and (\ref{mom}) (or, rather, Eqs. (\ref{hs}) and (\ref{ms})). {\it A priori,} at least, this seems to be a reasonable task. We have two free functions, $K_R$ and $K_l$, and we need to pick them so as to satisfy two scalar equations. However, we do not have a completely free choice. Given the metric, we can compute $M$ via Eq.(\ref{M}) and if we succeed in embedding this in a Schwarzschild solution of mass $m$, we must have that $m \ge M$. In addition to being a necessary condition, it is {\it almost} a sufficient condition. More precisely, we can prove: 

{\bf Proposition:} Given a spherical Riemannian three-geometry with finite $M$ as defined by Eq.(\ref{M}) and any (extended) vacuum Schwarzschild solution whose mass $m$ satisfies $m > M$, then a spacelike slice can be found in this spacetime which is isometric to the given three-geometry. 

{\bf Proof:} We start off with the mass expression, Eq.(\ref{m}), and write it as \begin{equation} K_R = \sqrt{{2m \over R^3} - {1 \over R^2}\left[1 - R'^2\right]}. \label{K_R} \end{equation} The quantity under the square root is positive definite since $m > M$ and so the equation makes sense. We can choose either the positive or the negative root. With either choice we get a well-defined function $K_R(l)$ which does not change sign. It may well be that if $R \rightarrow 0$ then $|K_R| \rightarrow \infty$. All this does is reflect the fact that the slice goes into the Schwarzschild singularity. 

Given that $K_R$ is bounded away from zero we now can solve the Hamiltonian constraint, Eq.(\ref{hs}), for $K_l$. More precisely, we rewrite Eq.(\ref{hs}) as 
\begin{equation} \label{hs1} 2K_R K_l +{2m \over R^3} - {2R'' \over R} = 0 \end{equation} 
and manipulate this to get 
\begin{equation} \label{K_l} K_l = {1 \over K_R}\left[{R'' \over R} - {m \over R^3}\right]. \end{equation} 
This is clearly well-defined and finite except possibly, again, as $R \rightarrow 0$. It is very straightforward to show that $K_R$ from Eq.(\ref{K_R}) and $K_l$ from Eq.(\ref{K_l}) satisfy the momentum constraint, Eq.(\ref{ms}). 

We have only shown that we can construct spherically symmetric initial data that satisfies the vacuum Einstein constraints. However, the Einstein evolution equations allow us to propagate this data so as to construct at least a patch of spacetime (which will be spherically symmetric) and which satisfies the vacuum Einstein equations. In turn, Birkhoff's theorem guarantees that this must be part of the Schwarzschild solution with given mass $m$. 

\section{A concrete example}
We wish to embed the `static' (moment-of-time-symmetry) slice from one Schwarzschild solution (of mass $m_1$) in another Schwarzschild solution (of mass $m_2$). We assume $m_2 > m_1$. The three-metric we are given, written in Schwarzschild coordinates is
\begin{equation}
ds^2 = {dR^2 \over 1 - {2m_1 \over R}} + R^2 d\Omega^2. \label{m_1}
\end{equation}
To convert to proper distance gauge we would need to integrate
\begin{equation}
l(r) = \int_{2m_1}^r {dR \over \sqrt{1 - {2m_1 \over R}}}
\end{equation}
but there is no real need to do so, all we really use is
\begin{equation}
R' = {dR \over dl} = \sqrt{1 - {2m_1 \over R}} \Rightarrow {R \over 2}\left[1 - R'^2\right] = m_1.
\end{equation}
When this is substituted into Eq.(\ref{K_R}) we get
\begin{equation}
K_R = -\sqrt{2(m_2 - m_1) \over R^3}.
\end{equation}
We choose the negative root to get the slice in the upper half plane.

We can use the Hamiltonian constraint, Eq.(\ref{hs}), remembering that the scalar curvature of the metric given by Eq.(\ref{m_1}) vanishes, to get
\begin{equation}
K_l = -{K_R \over 2} = \sqrt{m_2 - m_1 \over 2R^3}.
\end{equation}
It is easy to confirm that this choice of extrinsic curvature satisfies the momentum constraint, Eq(\ref{ms}).

We know that maximal slices which run from one end to the other of a Schwarzschild solution cannot approach the singularity closer than $R = 3m/2$ \cite{bom}. If we relax the maximal condition, and replace by the requirement that the three-scalar-curvature be nonnegative we get no such restriction, we can approach the singularity as closely as we wish.

If we let $m_1 \rightarrow 0$ we get the well-known flat slice of the Schwarzschild solution. Now $R \rightarrow 0$ so the flat slice runs into the singularity and the extrinsic curvature is given by
\begin{equation}
K_R = -2K_l = -\sqrt{2m_2 \over R^3}.
\end{equation}
We can also embed the `static' slices of the negative mass Schwarzschild solution in flat spacetime. These slices have intrinsic metric
\begin{equation}
ds^2 = {dR^2 \over 1 + {2m_1 \over R}} + R^2 d\Omega^2, \label{m_2}
\end{equation}
and extrinsic curvature
\begin{equation}
K_R = -2K_l = -\sqrt{2m_1 \over R^3}.
\end{equation}
This slice is regular everywhere except at the origin.

\section{The significance of \lowercase{$m$} $> M$} We seek the maximum of ${R \over 2}[1 - R'^2]$. If we wish to embed this slice in a negative mass Schwarzschild solution (or in flat spacetime) we need ${R \over 2}[1 - R'^2] < 0$ over the entire slice. This implies $R' \ge 1$. The areal radius starts off at $R = 0$ and monotonically increases, there cannot be a throat. Not only that but it must increase rapidly. It must look like a trumpet. This is completely in agreement with the requirement that the topology of slices embedded in the negative mass Schwarzschild must be $R^3$. 

In general, however, the maximum of ${R \over 2}[1 - R'^2]$ will be positive and therefore this slice can only be embedded in a positive mass Schwarzschild solution. For example, if the area is not monotonic then the function $R(l)$ has either a local maximum or minimum (or both). Thus the slice has at least one point where $R' = 0$, and if the areal radius equals $R_0$ at that point then  ${R \over 2}[1 - R'^2] = R_0/2$ there. Thus we have that $M \ge R_0/2$ and a necessary condition that this slice be embeddable in a Schwarzschild solution of mass $m$ is that $2m \ge R_0$. This means that if we have an extremum of the area we can only embed this slice in a positive mass Schwarzschild solution and also that all maxima and minima of the areal radius must occur inside the horizon. 

If we have a local maximum of $R(l)$, i.e., a point where $R' = 0, R'' \le 0$, this point is also a local maximum of ${R \over 2}[1 - R'^2]$. This is easy to show, all one needs to note is that both $R/2$ and $[1 - R'^2]$ have a maximum at that point. No equivalent statement can be made about a point where $R(l)$ is a local minimum. One could have a maximum or minimum of ${R \over 2}[1 - R'^2]$ at that point or nothing at all. It is also worth noting that the maximum of ${R \over 2}[1 - R'^2]$ need not occur at an extremum of $R(l)$. It may occur at a `large' value of $R$ where $R'$ may be small without ever going to zero. One can then choose an $m > M$ such that this point is outside the horizon. 

\section{The marginal case: what happens when $\lowercase{m} = M$} 
If we have a spherical slice in a Schwarzschild solution we know that $m \ge M = \max {R \over 2}[1 - R'^2]$. Conversely, we have shown that if we have a spherically symmetric Riemannian three metric we can embed it in a Schwarzschild solution with mass $m$ if $m > M$. In this section we would like to discuss the issue of when and if one can embed a given spherical slice in a Schwarzschild solution satisfying $m = M$. Very useful tools for analysing spherical spacetimes are the optical scalars, $\Theta_\pm$ (see e.g. \cite{gom,mom}). These are defined by \begin{equation} \label{theta} \Theta_\pm = {2 \over R}\left(R' \pm RK_R\right). \end{equation} These are the divergences of the future pointing and past pointing outward radial light rays from the isometry spheres. Both are defined so as to be positive on flat space and near infinity. Note that the definition of extrinsic curvature used agrees with Wald \cite{w} and not with Misner, Thorne, and Wheeler \cite{mtw}. 

The mass expression, Eq.(\ref{m}), can be rewritten as \begin{equation} {2m \over R} = 1 - {\Theta_+\Theta_-R^2 \over 4}. \end{equation} If $m \le 0$ then $\Theta_+$ and $\Theta_-$ must always remain positive. If $m > 0$ and if $2m/R > 1$, i.e., if we are inside the horizon, then one or other of the optical scalars must be negative. In the upper half of the extended Schwarzschild solution it is $\Theta_+$ which is negative and in the lower half it is $\Theta_-$ which is negative. Eq.(\ref{K_R}) is double valued because we can choose either sign when we take the square root. This means that we have two slices with the same intrinsic geometry. Given a solution to Eqs.(\ref{hs}, \ref{ms}) and multiply $K_R$ and $K_l$ by a minus sign we again have a solution with the same three-geometry. These two solutions  are just reflections of each other about the $t = 0$ plane. The slice with $K_R > 0$ crosses the horizon below $t = 0$ and either crashes into the past singularity or continues through while the slice with $K_R < 0$ tends to live in the upper half plane. 

Say we are given a spherical three-geometry with finite $M$ and try to embed this in a Schwarzschild spacetime with $m = M$. This means that we no longer, from Eq.(\ref{K_R}), get that $K_R > 0$. Rather we get that $K_R = 0$ at the point(s) where $M$ achieves its maximum. At such a point, since $\left({R \over 2}[1 - R'^2]\right) $ is at its maximum, we know that 
\begin{equation} 0 = {\partial \over \partial l}\left({R \over 2}[1 - R'^2]\right) = {R^2R' \over 4}\times {2 \over R^2}\left[1 - R'^2 - 2RR''\right]. \label{R} \end{equation} 
Therefore at the point(s) where $K_R = 0$ we are guaranteed that {\it either} $R' = 0$ or ${2 \over R^2}\left[1 - R'^2 - 2RR''\right] = 0$. From Eq.(\ref{hs}) it is clear that we need the second expression (which is nothing else but the three scalar curvature, ${\cal R}^{(3)}$, of the three-geometry) to vanish whenever $K_R = 0$. Hence we need to distinguish between the case where the maximum of ${R \over 2}[1 - R'^2]$ occurs at a point where $R' = 0$ (which is bad, except if simultaneously ${\cal R}^{(3)} = 0$ there, which cannot be guaranteed) and the case where the maximum occurs at a point where $R' \neq 0$ (which is good because ${\cal R}^{(3)} = 0$ there). This question is trivial in the case where $M \le 0$ because there cannot be any point(s) with $R' = 0$ in the geometry and so we automatically have that ${\cal R}^{(3)} = 0$ at the point where $M$ achieves its maximum. 

It is also clear that we  are unable to use Eq.(\ref{K_l}) directly to evaluate $K_l$ at the point(s) where $K_R = 0$. However, we can return to the momentum constraint Eq.(\ref{ms}) and rewrite it as \begin{equation} K_l = {RK_R' \over R'} + K_R. \label{K_l1} \end{equation} This shows us that $K_l$ is finite and well defined at the point(s) where $K_R = 0$ so long as $R' \neq 0$ at those points. This is consistent with Eq.(\ref{K_l}) because the term in square brackets in this equation is essentially ${\cal R}^{(3)}$ so Eq.(\ref{K_l}) becomes $K_l = 0/0$ and use of de l'Hospital's rule gives us a finite value. Therefore the only case we need to worry about is when the maximum of $M$ coincides with $R' = 0$. We cannot, in general, hope to solve the constraints with $m = M$ in this situation. Nevertheless, there are special cases.

It is clear that if there exists a point where $K_R = 0$ and $R' = 0$ then both optical scalars, $\Theta_{\pm}$, vanish simultaneously. In the Schwarzschild solution the only place where this can happen is at the bifurcation `point', where both horizons cross. Note again that $R' = 0$ is only possible with $m > 0$. If the given three-geometry has the property that that the maximum of ${R \over 2}[1 - R'^2]$ occurs at a point where $R' = 0$ and we try $m = M$, then clearly $K_R = 0$ and $R' = 0$ at this point; if this slice is to be embedded in this Schwarzschild solution it must pass through the bifurcation point. This places both local and global restrictions on the geometry of the slice. This point where $R' = 0$ must be a local minimum of the area, i.e., $R'' \le 0$. More than that, we require that the scalar curvature vanishes at that point, hence $2R'' = -1/R$. Further, it must be the global minimum. The area cannot oscillate. If we had a local maximum of $R$ then the value of ${R \over 2}[1 - R'^2]$ at this point would be larger than the value at the `throat', where it is supposed to be the maximum. Therefore we can only have one minimum. 

The behaviour of $K_R$ in the neighbourhood of the point where $K_R = 0$ is worth noting. Let us define $f(l) = {R \over 2}[1 - R'^2]$ and assume $f = f_0 = m$ at $l = l_0$, $f'_0 = 0$, and $f''_0 < 0$. We have $K_R^2 = {2m - 2f \over R^3}$. Hence we have $(K_R^2)_0 = 0$, $(K_R^2)'_0 = 0$, and $(K_R^2)''_0 = ({-2f'' \over R^3})_0 > 0$. Now make a Taylor expansion of $K_R^2$ around $l = l_0$ in terms of $x = l - l_0$. In general we will get $K_R^2(x) = A + B x + C x^2 + \dots$ with $A = B = 0$ and $C > 0$. Thus we get $K_R(x) = \sqrt{C} x + \dots $. Therefore $K_R$ will pass through zero at $l = l_0$ with nonzero slope and so must change sign. 

There is a further special case. The requirement that $f(l)$ be a maximum at $l = l_0$ only requires $f''_0 \le 0$. If $f''_0 = 0$ we would have $(K_R^2)_0 = 0$, $(K_R^2)'_0 = 0$, and $(K_R^2)''_0 = ({-2f'' \over R^3})_0 = 0$. Now the requirement that we have a maximum of $f$ forces $f'''_0 = 0$ and the first nontrivial derivative must be at fourth order. In this case the Taylor expansion of $K_R^2$ can only start at $x^4$, i.e. $K_R^2 = Dx^4 + \dots$ and $K_R = \sqrt{D} x^2 + \dots$. In this case $K_R = 0$ and $K_R' = 0$ at $l = l_0$. Following from this not only does $K_R$ vanish at $l = l_0$ but also, from Eq.(\ref{K_l1}), we have that $K_l = 0$ at $l_0$. 

To summarize: If the maximum of $ {R \over 2}[1 - R'^2]$ occurs at a point
where $R' \ne 0$ we can choose $m = M$. This condition of $R' \ne 0$ is trivially satisfied if $M < 0$. If the maximum occurs at a point where
$R' = 0$ we can only choose $m = M$ if this point becomes the bifurcation point. This places further requirements (both local and global) on the three-geometry.

\section{Spherical Superspace and the thick sandwich theorem}  Wheeler identified the configuration space of canonical general relativity as being the space of all spacelike three-geometries, which he called `superspace'. A trajectory of a solution of the Einstein equations in this configuration space corresponds to the sequence of Riemannian three-geometries generated by a foliation. This is not a unique curve, rather we have `a spray of geodesics' in the language of DeWitt. Every slicing of a spacetime generates a different sequence of three-geometries and by changing the slicing one changes the curve through superspace.  It is interesting to ask what fraction of superspace do all the curves corresponding to one given spacetime pass through. This set of three-geometries which is the union of all the solution curves is exactly the same as the set of all spacelike three-geometries that can be embedded in a given spacetime. 

 Let us define spherical superspace $(SS)$ as the space of all spherically symmetric spacelike three-geometries. Now consider a Schwarzschild solution $(S(m))$, which is defined by its mass $m$. There will be many spacelike spherical three-geometries that can be embedded in $S(m)$. We define $B_S(m)$ as the collection of all spherically symmetric spacelike three-geometries that can be embedded in a given $S(m)$. Obviously $B_S(m) \subset SS$. The analysis so far tells us a great deal about the relationship between $B_S(m)$ and $SS$. 

It is clear that any single $B_S(m)$ with given mass $m$ covers a large fraction of spherical superspace. We have a measure on spherical spacelike three-geometries given by $M = \max {R \over 2}\left[1 - R'^2\right]$ and we have shown that all three-geometries which satisfy $M < m$ belong to $B_S(m)$. We also know that any three-geometry which satisfies $M > m$ cannot belong to $B_S(m)$. Thus $B_S(m)$ is defined by a single algebraic condition. The only uncertainty is about those three-geometries on the boundary of $B_S(m)$ i.e., those metrics which satisfy $M = m$. If $m\le 0$ we know that $B_S(m)$ is closed; every geometry which satisfies $M = m$ can be embedded in the appropriate Schwarzschild solution. This is because $M \le 0 \Rightarrow R' \ne 0$ so, as discussed in the previous section, we have no difficulty in solving for $K_l$ at the point where $K_R = 0$. If $m > 0$ we have that $B_S(m)$ is neither open nor closed. We have that most of the geometries on the boundary can be embedded. However, we have a relatively small class of geometries, those for which the maximum of $M$ occurs at a local maximum of the area, i.e., a point where $R' = 0$, which cannot be embedded in the Schwarzschild solution with mass $m = M$. 

Further, we have a nested structure on spherical superspace. Given two Schwarzschild solutions with masses $m_1$ and $m_2$ with $m_1 < m_2$ then every spherical slice that can be embedded in the first solution can also be embedded in the second one. This means that $B_S(m_1) \subset B_S(m_2)$.  The foliation freedom in a given solution to the Einstein equations is represented by the fact that one can choose an arbitrary lapse (the shift freedom can be ignored because we are considering the geometries rather than three-metrics). Starting with spherical data and maintaining the spherical symmetry means that we must have a spherical lapse. Therefore the foliation choice is represented by one single spherical function. However, as we have seen in Eq.(\ref{g}), the freedom in spherical three-geometries is also represented by one single spherical function. Thus it should not come as too much of a surprise that any one Schwarzschild solution should cover so much of spherical superspace. 

Since this kind of counting argument works so well in the spherical case, it is interesting to try it in more general situations. Let us define axial superspace as the set of all axially symmetric spacelike geometries. The foliation freedom of axially symmetric slicings of an axially symmetric spacetime is represented by a single axially symmetric function. However, to give a general axially symmetric three-geometry (say written in the Brill conformal gauge \cite{b}) we have to give {\it two} axially symmetric functions. Therefore we expect that a given axially symmetric spacetime will visit only a subset of axially symmetric superspace, corresponding more-or-less to the square root of the whole. In the general, unsymmetric, case the foliation freedom is represented by a single arbitrary function while the general three-geometry is represented by three arbitrary functions.

Spherical geometries are also useful in that they offer a simple counterexample  to a longstanding idea in canonical gravity, the thick-sandwich approach.
 Wheeler introduced the concept of the `thick sandwich theorem'. The plan was to choose an initial and final point in configuration space (in this case a pair of three-geometries) and that the equations of motion (the dynamical part of Einstein equations) would give a natural path joining these two points. In other words to try and find a unique four manifold, satisfying the Einstein equations, which fills in between the two given three-geometries. The analysis in this article shows how ill posed this `thick sandwich' concept is.

 Choose any two spherical three-geometries. Evaluate $M$ for each geometry. Let them be $M_1$ and $M_2$ respectively. Pick any Schwarzschild solution with mass $m$ which satisfies $m > \max(M_1, M_2)$. Both of the given three-geometries can simultaneously be embedded in the given Schwarzschild solution. This generates a `filling' (which satisfies the Einstein equations) between the given slices which is highly nonunique. Further, it is impossible to lift this ambiguity by demanding that the given three-geometries are, with respect to any measure, close to one another. If we find any Schwarzschild solution that interpolates between them, all Schwarzschild solutions with larger mass will also do the job. 

\acknowledgments K.R.\ acknowledges the support of the TEMPUS project JEP-12249-97. We wish to thank Edward Malec for much help.

\end{document}